\documentclass{article}

\usepackage{arxiv}

\usepackage[utf8]{inputenc}
\usepackage[T1]{fontenc}
\usepackage{lmodern}
\usepackage{graphicx}
\usepackage[figurename=Fig.,labelfont=bf,labelsep=period]{caption}
\usepackage{subcaption}
\usepackage{amsmath}
\usepackage{newtxtext,newtxmath}
\usepackage[colorlinks=true,citecolor=red,linkcolor=black]{hyperref}
\title{Quantifying Resilience via Multi-Scale Feedback Loops in Water Distribution Networks}

\author{
Alessio~Pagani\\
The Alan Turing Institute, London\\
\texttt{apagani@turing.ac.uk} \\
\And
Fanlin~Meng \\
University of Exeter, Exeter\\
\And
Guangtao~Fu \\
University of Exeter, Exeter\\
The Alan Turing Institute, London\\
\And
Mirco~Musolesi \\
University College London, London\\
The Alan Turing Institute, London\\
\And
Weisi~Guo \\
University of Warwick, Coventry \\
The Alan Turing Institute, London\\
\\
}

\date{}

\begin{document}

\maketitle

\begin{abstract}
Water distribution networks (WDNs) are one of the most important man-made infrastructures. Resilience, the ability to respond to disturbances and recover to a desirable state, is of vital importance to our society. There is increasing evidence that the resilience of networked infrastructures with dynamic signals depends on their network topological properties. Whilst existing literature has examined a variety of complex network centrality and spectral properties, very little attention has been given to understand multi-scale flow-based feedback loops and its impact on overall stability. 

Here, resilience arising from multi-scale feedback loops is inferred using a proxy measure called \textit{trophic coherence}. This is hypothesised to have a strong impact on both the pressure deficit and the water age. Our results show that trophic coherence is positively correlated with the time to recovery (0.62 to 0.52), while it is negatively correlated with the diffusion of a disruption (-0.66 to -0.52). Finally, apply random forest analysis is used to combine resilience measures, showing that the new resilience ensemble provides a more accurate measure for networked resilience.
\end{abstract}

\section{Introduction}
Water distribution systems (WDSs) and Water distribution networks (WDNs) are lifeline systems of urban cities for the safe and reliable supply of drinking water \cite{vorosmarty10,Mekonnene1500323,short-of-water}. It is critical to build resilient WDSs \cite{Greco12,Fu16,Herrera18} so that desirable water service can be efficiently restored once a failure (e.g., pipe burst \cite{farley05}, contamination \cite{Pye713}) occurs in a system. 

Water distribution is under increased stress from rising human demand and drought due to climate change. In the UK, it is expected that 4,000 Mega litres/day (26\% increase) of extra water is needed in the near future \cite{NIC18}. Failure to respond to stressors can lead to a £40bn cost in emergency response. It is expected that improving the resilience of water distribution systems will cost £21bn, and the primary focus areas include reducing leakage and demand, as well as improving demand management and resilience to stressors (present and future). This is part of wider resilience frameworks (e.g., City Resilience Index - Arup \& Rockefeller Foundation, and Ofwat Towards Resilience) \cite{Welsh17}. According to the guaranteed standards scheme (GSS) \cite{gss-water} in the UK, water companies will be penalised if water supply is not restored within 12 hours the supply was cut off, or pressure falls below a minimum threshold limit on two occasions, each occasion lasting more than one hour, within a 28-day period. To meet the growing customer demands/regulatory requirements, it is essential to incorporate resilience in the design and operation of WDSs.

Here, resilience is defined as the ability of the network to recover to a predefined context-dependent desirable operating state, after suffering an extreme disturbance \cite{doi:10.1177/0956247807076726}. In the context of WDNs, several disturbances are possible and, accordingly, resilience can be measured from different perspectives: for example, one aspect is the water deficit and the water age increase caused by pipe bursts, another complementary aspect is for a contamination to be flushed out.

Methods have been reported on resilience and robustness assessment of WDSs, such as the Global Resilience Analysis based on repetitive stress and strain tests under different demand scenarios \cite{MUGUME201515,Fu16}, or entropy-based metrics that measure surplus energy in a WDS \cite{TODINI2000115}. For example, random or targeted (e.g., high degree) closure of junctions can cause a drop of WDS performance. Mapping performance to macroscale (e.g., the degree distribution \cite{Jeffrey11}) and mesoscale (e.g., core-periphery ratio \cite{Ma15}) network properties is standard practice in analysing the robustness of dynamic complex networks \cite{Jeffrey10,Greco12}. Whilst these methods can provide detailed resilience performance results, however, some of the research are built on extended period simulations hence require hydraulic/water quality models of WDSs.

Other methods seek to infer resilience knowledge from complex network property-based indicators. For example, surrogate metrics such as network connectivity, efficiency, modularity, bridgeness, and centrality are found to be strongly correlated with detailed resilience performance \cite{MENG2018376}. However, as reviewed in Table \ref{table1} (other similar reviews found at \cite{Giudicianni18}), most of the network science measures or approaches are not relevant to WDN dynamics or capture only a small part of the wider complex system.
For example, algebraic connectivity (second smallest eigenvalue of the Laplacian matrix \cite{Archetti15}) is closely related to robustness and phase synchronisation (e.g., for Kuramoto dynamics), which is not the underlying dynamic for WDNs.
Moreover, topological characteristics may not be fully representative of hydraulic performances and pipe failure impacts \cite{pagano19}.
Other approaches include partitioning the WDN into smaller sectors to improve management and data collection \cite{Nardo18C}, but these approaches do not give explicit relations between stability and network structure that includes dynamic flows.

In recent years, simplistic analytical mappings between explicit network dynamics that map complex network topology with local low-dimensional dynamics have been progressing from averaged dynamic estimation \cite{Gao16} to node-level precise estimators \cite{Moutsinas18}. However, their restriction to simplistic ODEs and homogeneous functionality means they cannot be effectively applied to complex WDSs.

Combined together, the aforementioned methods indicate a limitation in a deeper insight of WDS resilience assessment, as numerical analysis yields no tractable insight and an encompassing mathematical is absent. Despite the correlation with several network metrics \cite{MENG2018376}, current complex network analysis do not consider a key flow measure, namely the number of feedback loops \cite{Creaco15}. 

Resilience of directed networks with flow dynamics has gained increasing attention recently. When networks are modelled as a linear time invariant (LTI) system with a defined input and output \cite{Proakis:2006:DSP:1212134}, the dynamic response stability (resilience) is defined by the location of roots of its transfer function (e.g., stable if all in the negative domain). In such a case, absence of feedback loops ensures stability and the presence of feedback loops will risk instability, especially when there is uncertainty in the system. When non-LTI dynamics connected by an $N$-node complex network with $\sim N^{2}$ input output combinations are considered, a general transfer function cannot be easily defined. As such, alternative network stability metrics have been proposed to measure the number of multi-scale feedback loops. One such popular measure \textit{trophic coherence} is proposed by Johnson et al. \cite{Johnson14,Johnson17}, whereby it was shown that "a maximally coherent network with constant interaction strengths will always be linearly stable", and that it is a better proxy for linear stability than size or complexity. \textit{Trophic coherence} defines how much a directed graph is hierarchically coherent. The quantification of multi-scale feedback loops maps well to fractal properties found in existing WDNs \cite{Nardo17}. The rationale is that more hierarchically coherent systems have fewer feedback loops and are less likely to suffer from cascade effects. 

\begin{table}[t]\caption{Summary of Topological Resilience Metrics for Complex Networked Systems} \label{table1}
\begin{center}
\begin{tabular}{ccc}
\hline  Metric or Approach                          &Concept                    &Deficiency                     \\
\hline
        Mean degree \cite{networkx-paper}           &Avg. connectivity          & Biased by extreme values      \\
        Algebraic connectivity \cite{MENG2018376}   &Ease of synchronisation    & Incorrect dynamic for WDNs    \\
        Spectral gap \cite{Archetti15}              &Disconnectedness           & No dynamics considered        \\
        Eccentricity diameter \cite{MENG2018376}    &Max. disconnectedness      & Biased by periphery junctions \\
        Betweenness \cite{Simone18}                 &Bridging nodes between DMAs& Biased by bridge nodes        \\
        Efficiency \cite{MENG2018376}               &Ease of energy exchange    & No dynamics considered        \\
        Modularity \cite{Nardo18C}                  &No. partitions (DMAs)      & No dynamics considered        \\
        Core size \cite{Ma15}                       &Core reinforcement         & No dynamics considered        \\
        Node removal \cite{Jeffrey10}               &Cascade failure            & Incorrect dynamic for WDNs    \\
        Spectral gap \cite{MENG2018376}             &Eigenvalue distribution    & No dynamics considered        \\
        Trophic coherence \cite{Johnson14}          &No. feedback loops         & Lack of time lag              \\
\hline
\end{tabular}
\end{center}
\end{table}

In this paper is introduced a parameter for quantifying the resilience of weighted directed networks measuring their \textit{trophic coherence}. Trophic coherence, which is based on the concept of trophic levels in ecological systems, can be used as a proxy for measuring the stability of multi-scale feedback loops and motifs on large complex networks.
It has already been related to several structural and dynamical properties of directed networks.
Trophic coherence has been used to measure the stability/resilience on food web \cite{Johnson14}, epidemics \cite{klaise16}, transportation systems \cite{Pagani19}. However, no studies have been reported to examine its relationship with resilience performance of WDSs. 

The aim of this paper is to propose trophic coherence as a proxy measure of resilience performance of WDNs. Then, it will be demonstrated a strong correlation between the proposed coherence measure and some known resilience performance indicators using EPANET simulations of WDNs with different size, topological features, demand cycles, and flow dynamics. Different disruption scenarios including pipe burst and chemical contamination are simulated for analysis. Finally, random forest analysis is used to create a resilience ensemble metric, leveraging on both our proposed network feedback metric, as well as other state-of-the-art metrics.

\section{Methods}
Resilience assesses the dynamic system performance of networks under stress conditions, measuring the strength of the disruption and the time to recover.
In this study, resilience is measured using a proxy measure, the \textit{trophic coherence}. Trophic coherence groups in one value the quantity of loops and motifs, recognised as elements that interfere with the resilience of the WDNs.


\subsection{Measuring Resilience in WDNs}
Resilience can be measured from different perspectives, highlighting different aspects and properties of the WDNs. In this work, the resilience of the WDNs is measured stressing the network with different disruptions. Two stress test conditions are used: one revolves around a \textit{physical disruption} of the network (i.e., a pipe burst or a junction closure), the other concerns \textit{water contamination} with a chemical injected in a specific junction. For the former test, the resilience measures considered are: the number of junctions that suffer a demand deficit, the percentage of population affected by the breakage and the time required to restore the optimal conditions in terms of water age. For the latter condition, the resilience is computed as the time required by the WDN to expel the injected chemical, the extent of the chemical and the population contaminated.

\subsubsection{Resilience Measures}
\label{sec:resilience_measures}

\paragraph{Water Demand Deficit.}
A water quality simulation is used to compute demand at every junction over time. Firstly, the water demand is measured at each junction in standard condition, then, a breakage is simulated by closing a junction and the demand is measured again. The junctions downstream the breakage have to collect the water required to satisfy their demand from other water sources. If the other sources can not provide enough water or there are not other sources, these junctions experience a water demand deficit.
The water deficit is computed with two measures: (I) the number of junctions suffering demand deficit (e.g., junction receiving less that 25\% of required water demand) and (II) the percentage of population affected by the closure. The population connected to a junction is affected when the quantity of water they receive is below a given threshold (e.g., less that 25\% of required water demand). The process is repeated simulating a closure in every possible junction in the WDN and the average of the results is used to measure the resilience of a WDN.

\begin{figure}[!ht]
    \centering
    \begin{subfigure}[b]{0.48\textwidth}
      \includegraphics[width=\textwidth]{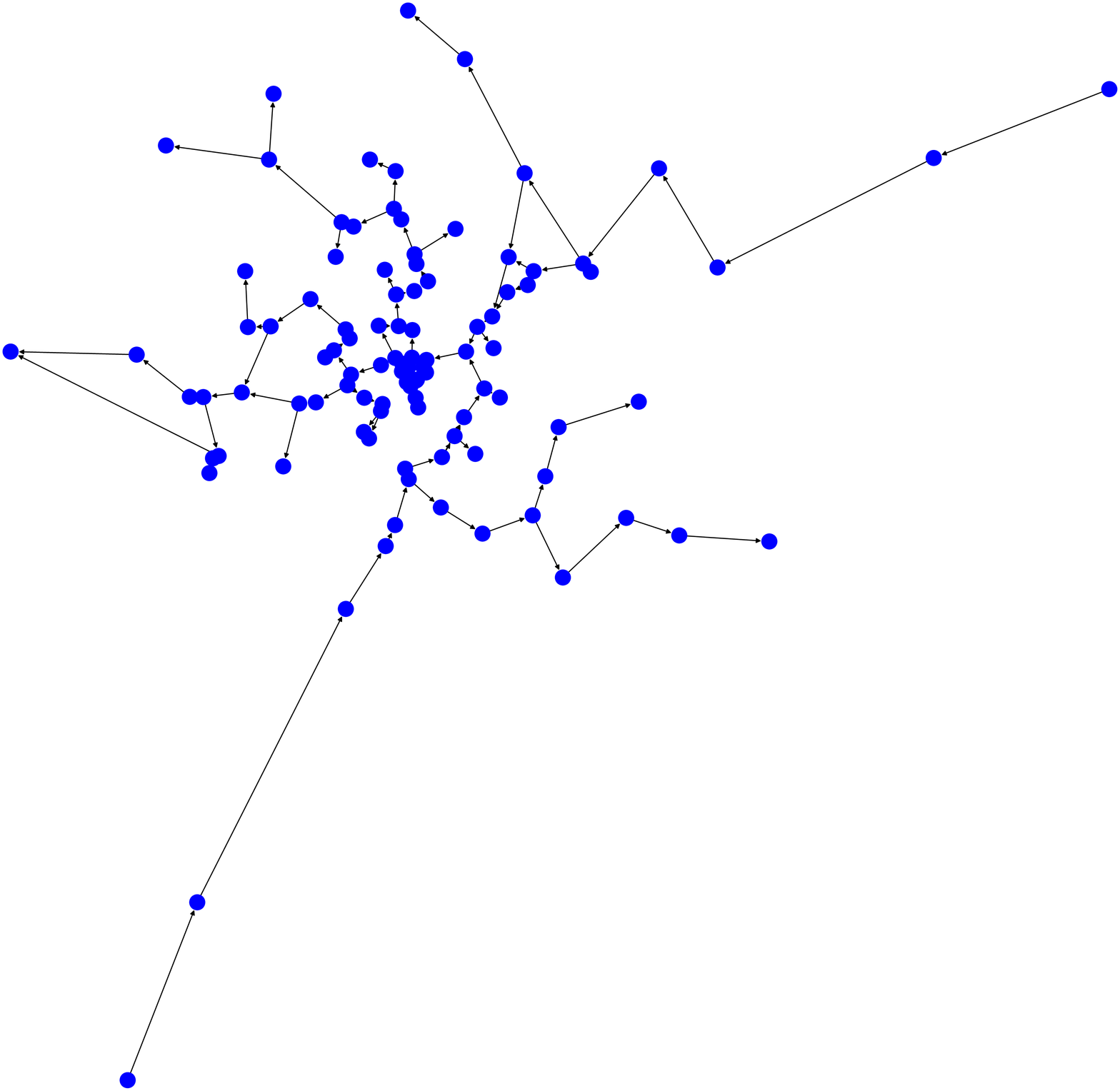}
      \caption{Network with optimal demand in each junction.}
      \label{fig:deficit-seq1}
    \end{subfigure}
    \hfill
    \begin{subfigure}[b]{0.48\textwidth}
      \includegraphics[width=\textwidth]{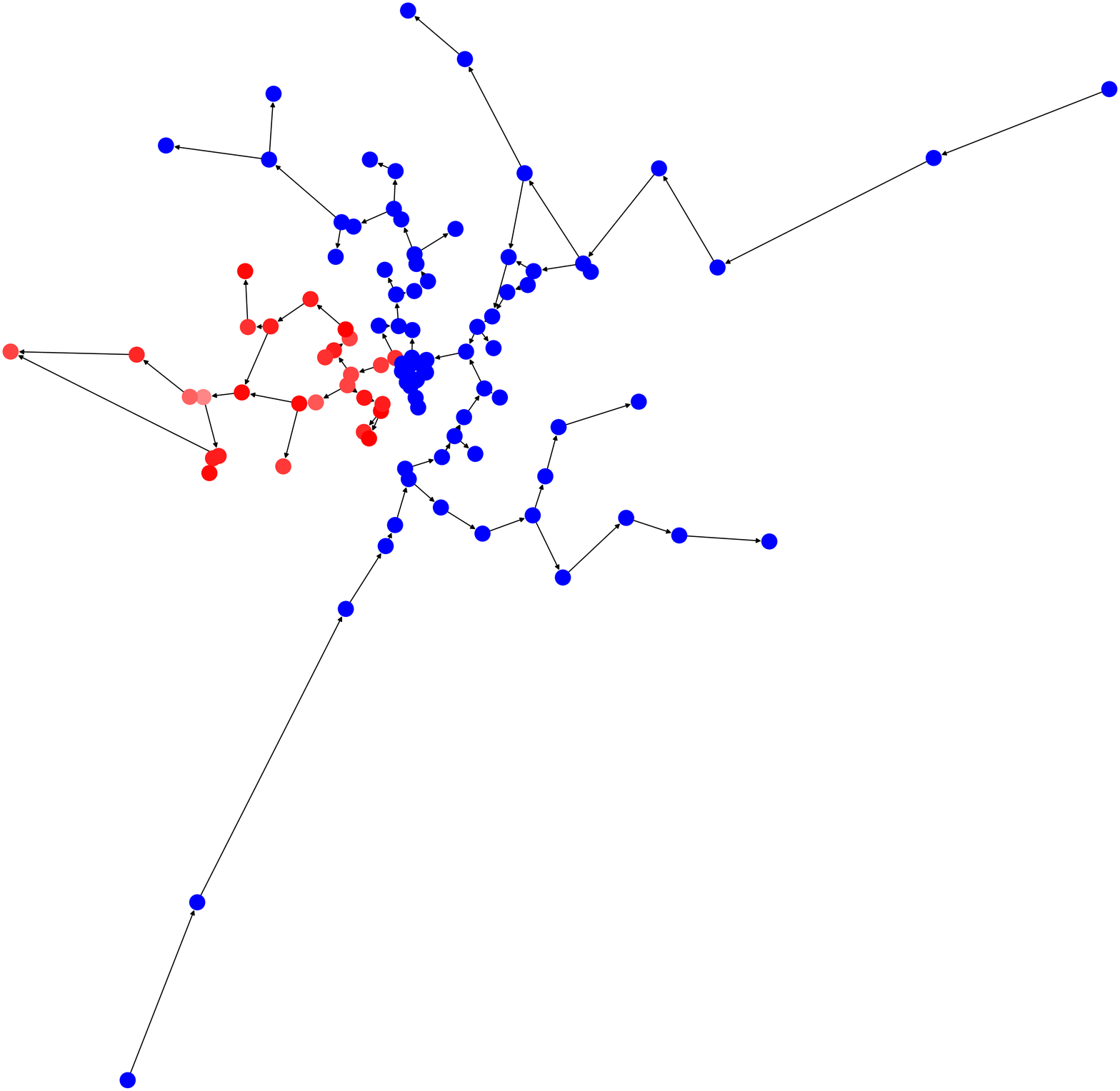}
      \caption{Junctions suffering pressure demand (in red) caused by a junction breakage.}
      \label{fig:deficit-seq2}
    \end{subfigure}
    \hfill
    \caption{Demand deficit.}
\end{figure}

An example of network in standard conditions, which delivers the required waterdat to each junction is shown in Figure \ref{fig:deficit-seq1}. The same network, after a junction closure is shown in Figure \ref{fig:deficit-seq2}: the junctions downstream the closure are isolated and do not receive water (red junctions).

\paragraph{Water Age.}
A water quality simulation is used to compute \textit{water age} at every junction. The water age at a junction is the time elapsed since the water currently in that junction entered the WDN from a reservoir or a tank.
The water age is measured at each junction in standard condition, then a breakage is simulated by closing a junction for a predefined amount of time and then reopening it. The water age at the junctions behave differently, according to the position of the junctions in respect of the closed one: the water age does not change in the junctions that are in a part of the network that is independent of the closure, it increases in the junctions that can still be reached, but with an alternative (longer) paths, it decreases to zero in the junctions that are isolated from reservoirs and tanks.
Eventually, when the junction is reopened, the water age gradually returns to standard conditions. The time needed to restore the standard conditions (i.e., the water age returns the same as before the disruption) is measured.
The process is repeated simulating a closure in every possible junction in the WDN and the average of the results is used to measure the resilience of a WDN.

\paragraph{Chemical Contamination.}
A water quality simulation is used to compute chemical concentration at each junction after the injection of a chemical in a specific point. The chemical is injected in a junction for a predefined amount of time. The chemical spreads in the WDN and it is expelled by demand junctions. Three measures are used to evaluate the impact of a chemical on a WDN: (I) the time required to clean the pipes and the junctions from all the chemical, (II) the maximum extent of the chemical in the pipes, (III) the percentage of population affected by the chemical.
The process is repeated simulating a chemical injection for every possible junction in the WDN and the average of the results is used to measure the resilience of a WDN.

\begin{figure}[!ht]
    \centering
    \begin{subfigure}[b]{0.48\textwidth}
      \centering
      \includegraphics[width=\textwidth]{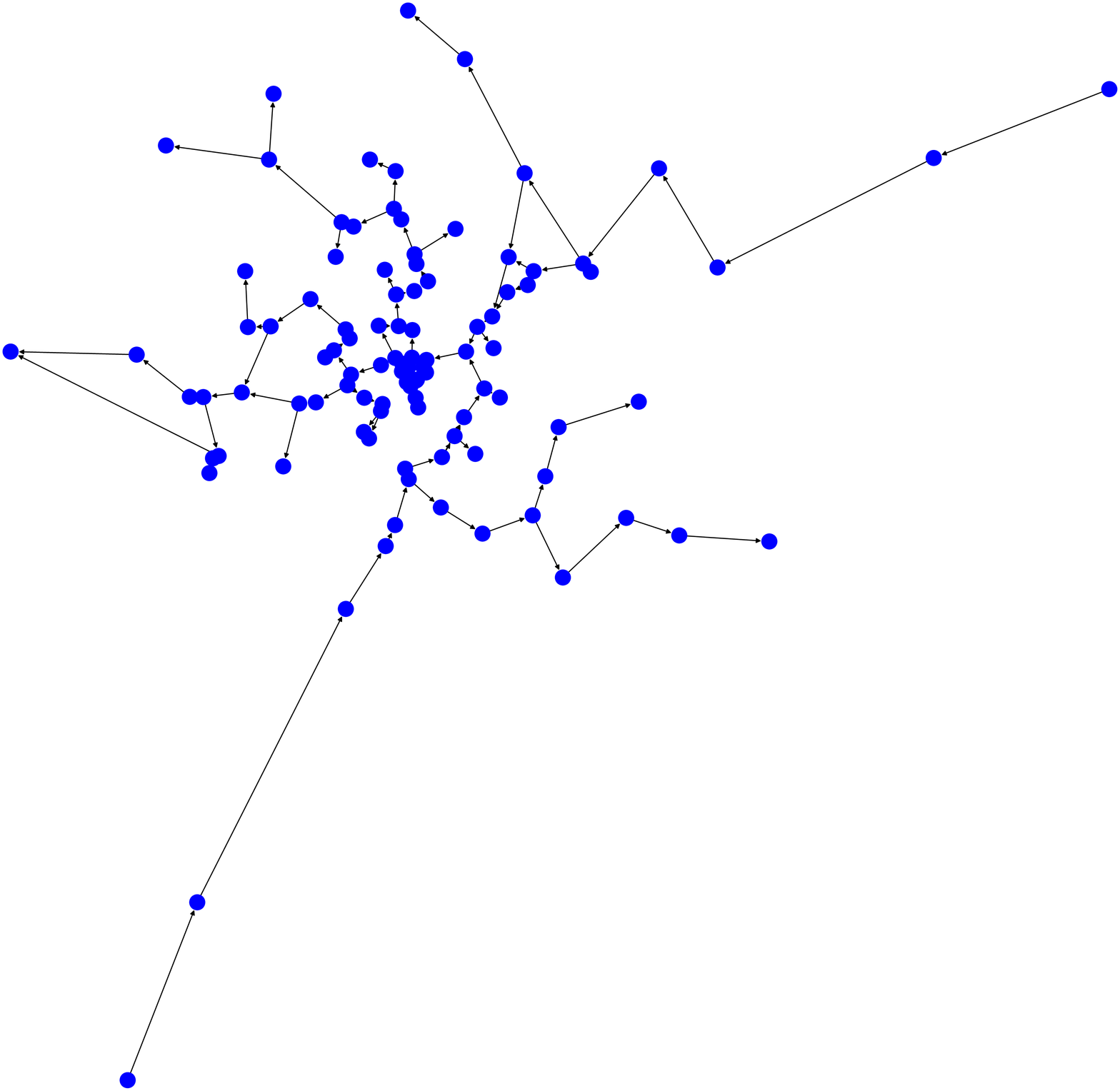}
      \caption{Network in standard conditions.}
      \label{fig:chemical-seq1}
    \end{subfigure}
    \hfill
    \begin{subfigure}[b]{0.48\textwidth}
      \centering
      \includegraphics[width=\textwidth]{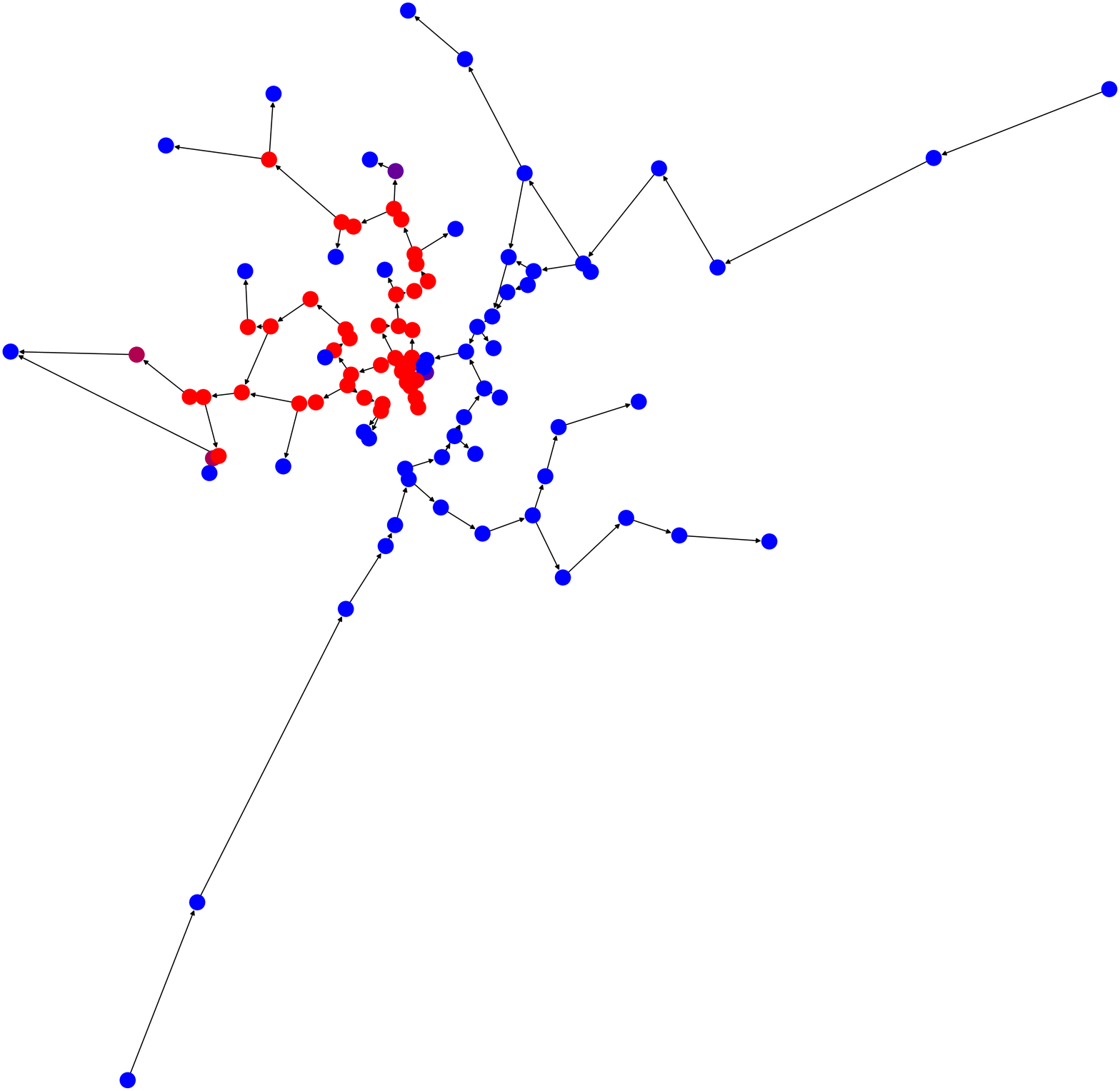}
      \caption{Chemical injection, the chemical spreads in the WDN (the affected junctions in red).}
      \label{fig:chemical-seq2}
    \end{subfigure}
    \hfill
    \begin{subfigure}[b]{0.48\textwidth}
      \includegraphics[width=\textwidth]{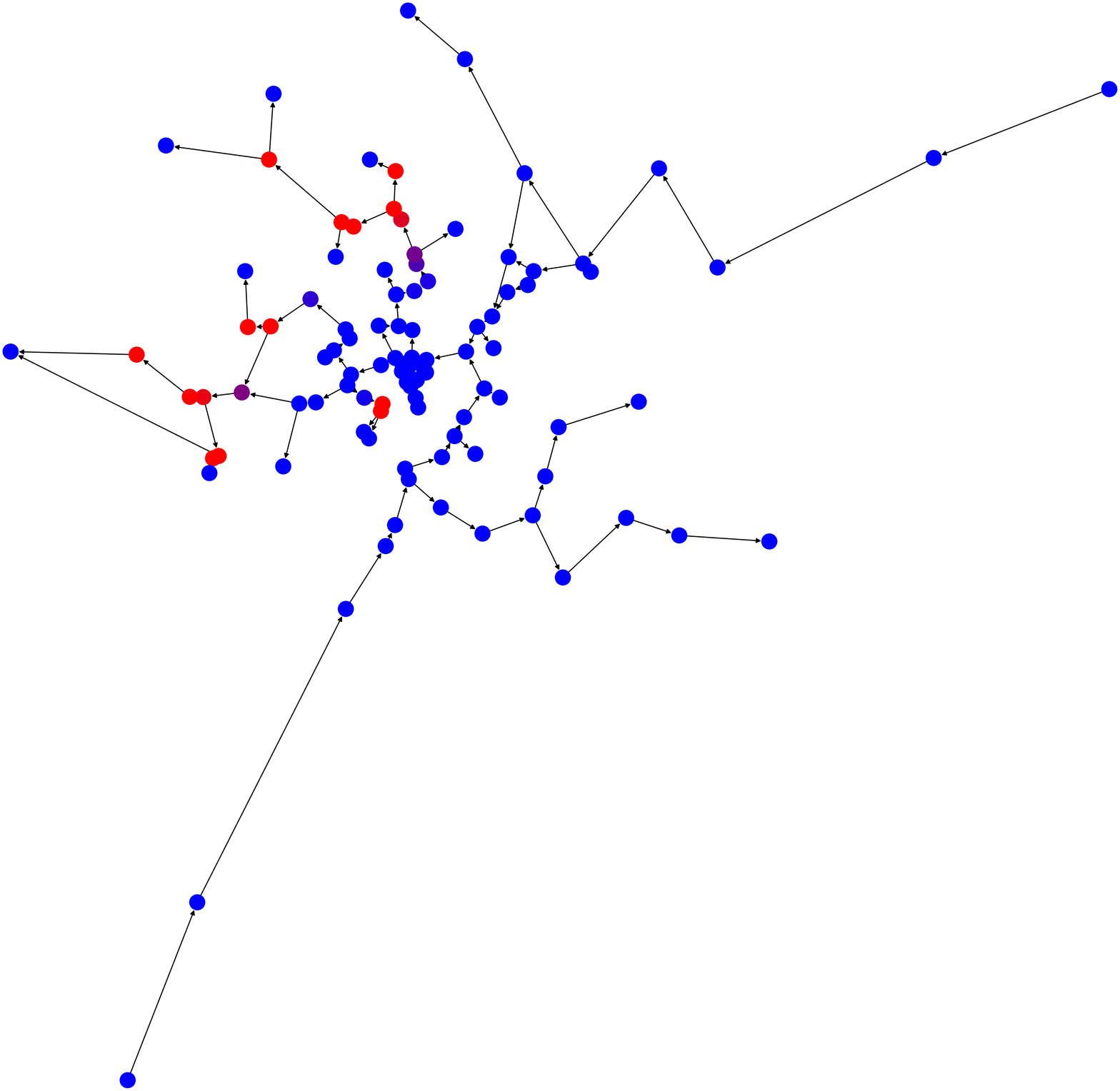}
      \caption{The chemical reaches the junctions downstream the injection (the affected junctions in red).}
      \label{fig:chemical-seq3}
    \end{subfigure}
    \hfill
    \begin{subfigure}[b]{0.48\textwidth}
      \includegraphics[width=\textwidth]{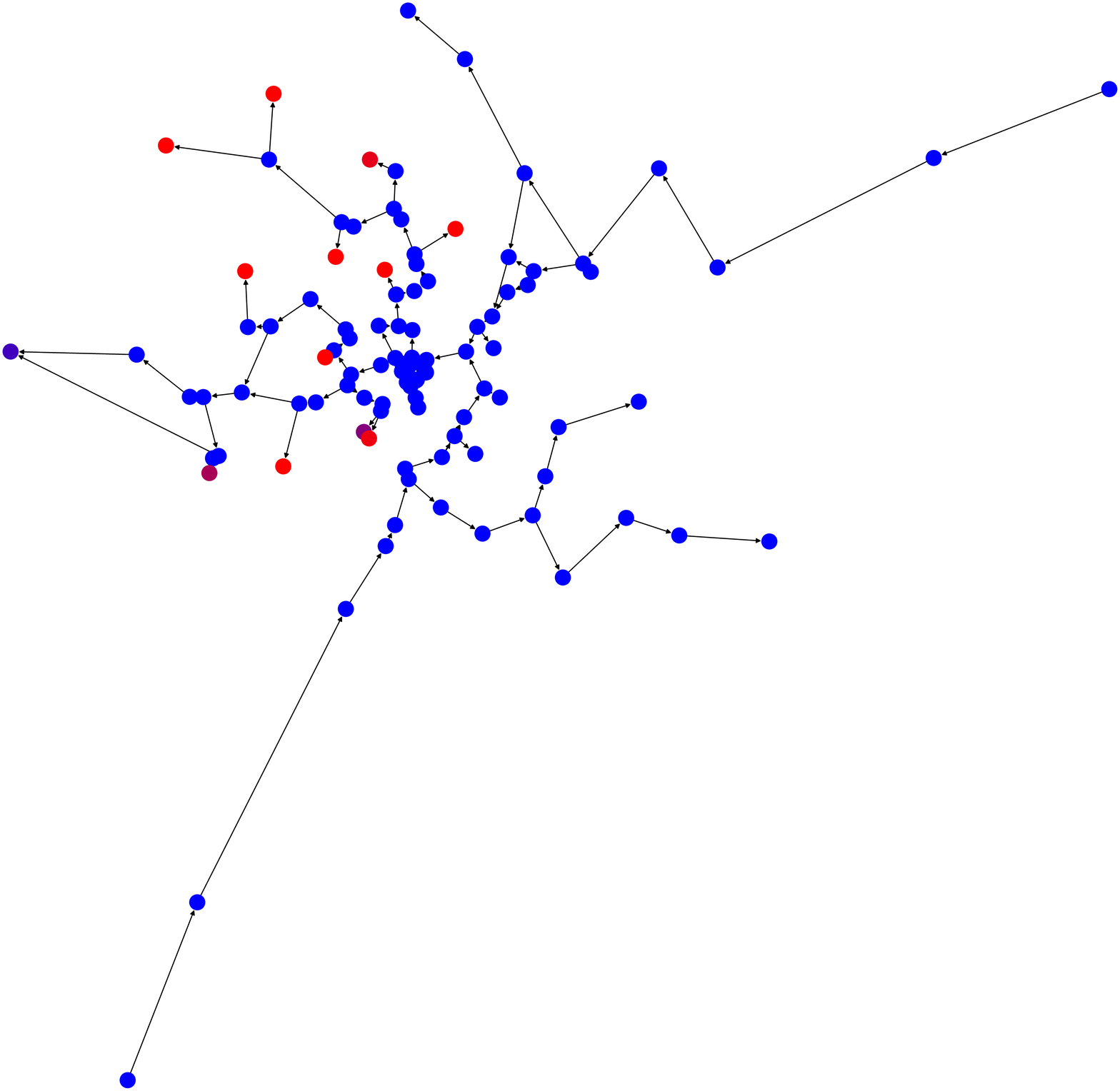}
      \caption{The chemical spreads in all the junctions downstream the injection (the affected junctions in red).}
      \label{fig:chemical-seq4}
    \end{subfigure}
    \hfill
    \caption{Chemical injection.}
\end{figure}

An example of chemical diffusion in the pipes and junctions over time is shown in Figure \ref{fig:chemical-seq1}, \ref{fig:chemical-seq2}, \ref{fig:chemical-seq3} and \ref{fig:chemical-seq4}.

\subsection{Trophic Coherence in WDNs}
\label{subsec:trCoherAndResilience}
Trophic coherence is based on the concept of trophic levels used mainly in ecology \cite{Johnson14}, but which can be defined for directed networks in general.

The trophic level of a node $i$, called $s_i$, is defined as the average trophic level of its in-neighbours, plus $1$:
\begin{equation}
s_i=1+\frac{1}{k_i^{\text{in}}}\sum_j a_{ij} s_j,
\label{eq_s_def}
\end{equation}
where $a_{ij}$ is an element of the adjacency matrix $A$ of the graph and $k_i^{\text{in}} = \sum_j a_{ij}$ is the number of in-neighbours (in degree) of the node $i$. Basal nodes $k_i^{\text{in}} = 0$ have trophic level $s_i = 1$ by convention.

By solving the system of equations (\ref{eq_s_def}), it is always possible to assign a unique trophic level to each node as long as there is a least one basal node, and every node is on a directed path which includes a basal node. 
In WDNs the basal nodes are the junctions that pump in water in the system (i.e., reservoirs and tanks), while the trophic level of each junction is the average level of all the junctions from which it receives water plus 1. For this reason, junctions near reservoirs and tanks will have lower trophic level than those far away from them.

Each pipe (edge) has an associated trophic difference $x_{ij}$:
\begin{equation}
x_{ij} = s_i-s_j,
\label{eq_tr_diff_def}
\end{equation}

The distribution of trophic differences, $p(x)$, always has mean 1, and the more trophically coherent a network is, the smaller the variance of this distribution. The trophic coherence can be measured with the \textit{incoherence parameter q}, which is simply the standard deviation of $p(x)$:\\
\begin{equation}
\label{eq:trophicLevels}
q = \sqrt{ \displaystyle\frac{1}{L}\displaystyle\sum_{ij} a_{ij} x_{ij}^2 - 1}
\end{equation}
where:
\newline
$L = \displaystyle\sum_{ij} a_{ij}$ is the number of pipes (edges) between the junctions (nodes) in the WDN. A perfectly coherent network will have $q = 0$, while a $q$ greater than $0$ indicates less coherent networks.

\subsubsection{Comparison with Null Model}
\label{subsubsec:basal_ensamble}
The degree to which empirical networks are trophically coherent (or incoherent) can be investigated by comparison with a null model. The \textit{basal ensemble expectation} $\tilde {q}$ can be considered a good approximation for finite random networks \cite{Johnson17,Pagani19}.
This parameter is used as a null model to compare the incoherence parameter of our empirical networks.

The basal ensemble expectation for the incoherence parameter is:
\begin{equation}
\label{eq:basal_ensamble}
\tilde {q} = \sqrt{ \displaystyle\frac{L}{L_B} - 1}
\end{equation}
where:
\newline
$L$ = number of edges in the network.\\
$L_b$ = number of edges connected to basal nodes.\\
\newline
The ratio $\displaystyle q/{\tilde {q}}$ is used to analyse the coherence of the network: a value close to 1 shows a network with a trophic coherence similar to a random expectation. Values lower than 1 reveal coherent networks, while values greater than 1 incoherent ones.


\subsubsection{Example}
\begin{figure}[ht]
    \centering
    \begin{subfigure}[b]{0.48\textwidth}
      \includegraphics[width=\textwidth]{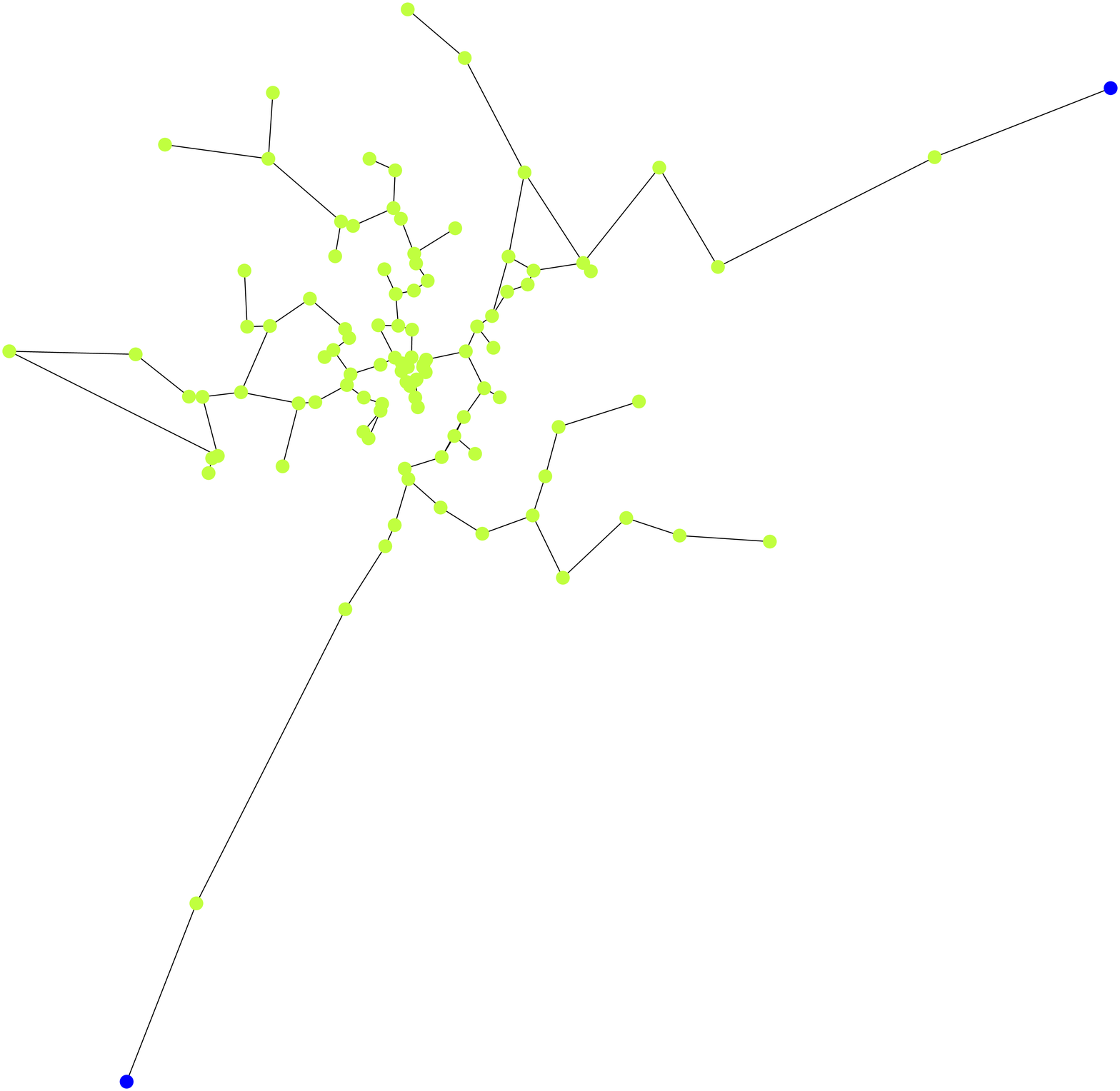}
      \caption{Example of WDN, reservoirs and tanks are highlighted in blue.}
      \label{fig:wdn-reservoir}
    \end{subfigure}
    \hfill
    \begin{subfigure}[b]{0.48\textwidth}
      \includegraphics[width=\textwidth]{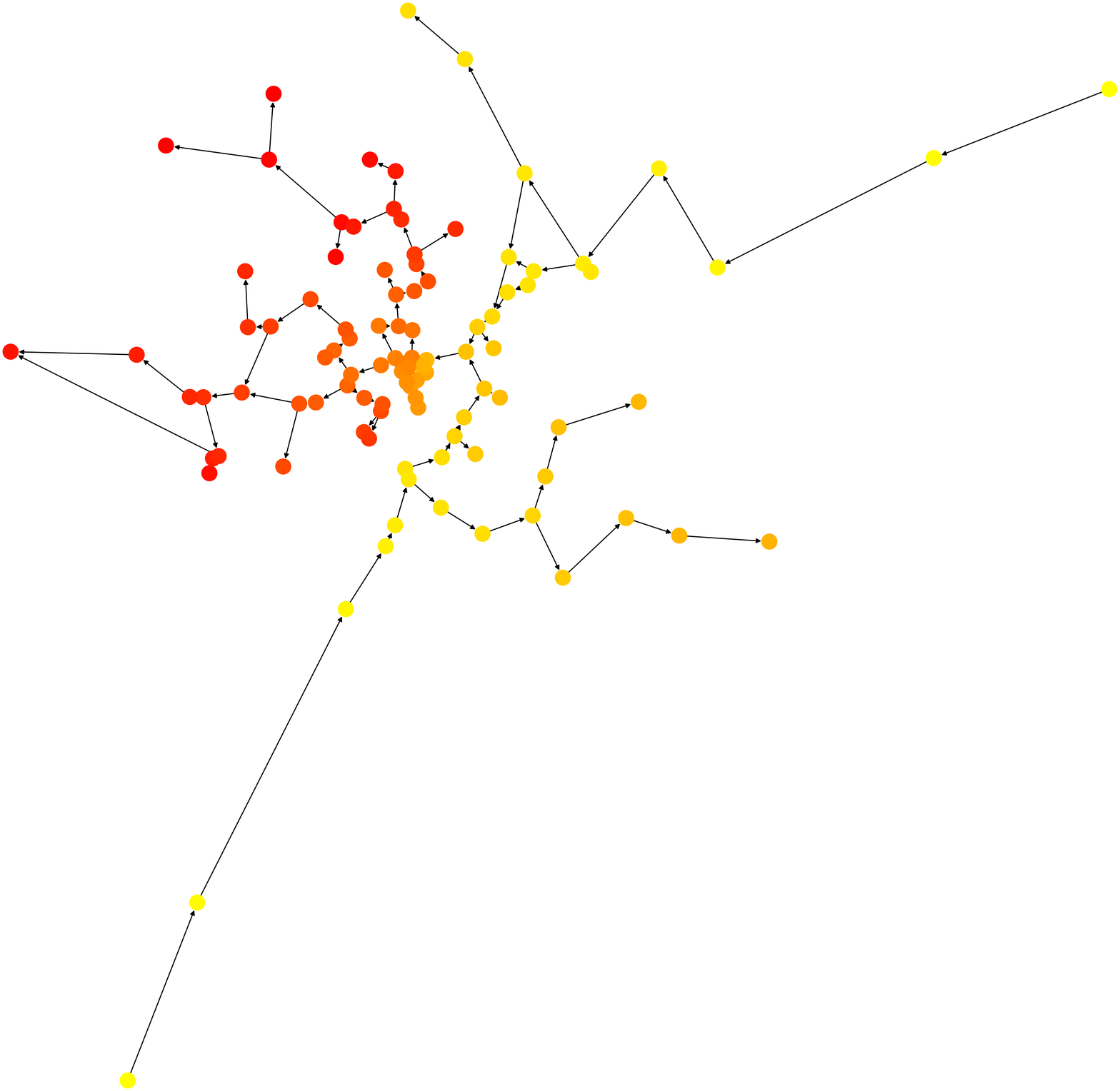}
      \caption{Trophic level of the junctions, from yellow (low trophic level) to red (high trophic level).}
      \label{fig:wdn-trophic_level}
    \end{subfigure}
    \hfill
    \caption{Trophic level of the junctions in a WDN.}
    \label{fig:wdn-example}
\end{figure}

An example is provided in Figure \ref{fig:wdn-example}: reservoirs, in yellow, have trophic level 1. The trophic level of the other junctions is represented with a colour from yellow (low trophic level) to red (high trophic level). 

\subsection{Water Distribution Networks Simulation}
The simulations in this work are conducted using the python package Water Network Tool for Resilience (WNTR) based on EPANET2 \cite{EPANET2}, which is capable of performing extended-period simulation of hydraulic and water-quality behaviour within pressurised pipe networks. The simulations are performed on clouding computing platform Microsoft Azure \cite{azure} for accelerated simulation performance.

The WNTR simulation engine is used to simulate the dynamics and the water flows of the networks (described in Section \ref{dataset}). All the experiments hereafter described are executed using the WNTR simulation engine and running the hydraulic simulation for 24 hours. The disruption scenarios tested could lead to low pressure conditions, for this reason the simulations are executed in pressure dependent demand scenarios (i.e., the actual delivered water depends on the pressure) thus consumers do not always receive their requested demand.

The experiments are firstly conducted with constant water demand at the junctions, that is, the water demand in each junction is predefined and does not change over time. Next, the experiments are repeated using variable demand patterns: the water demand changes over time simulating the variation in demand over a day, with peak and off-peak times. To increase the cascade effects, the disruptions (physical and chemical) are simulated during the peak time, where there is the major water demand.

\subsubsection{Baseline}
For each experiment, a simulation with no disruptions, hereafter called simulation in standard conditions, is used as baseline for the resilience measurement. The population served by each junction is computed according to the standard USEPA15 \cite{compute_population}.

\section{Dataset}
\label{dataset}
WDN models are used to represent network topology, water consumption, and control rules. Their main components are  the reservoir and tanks that provide water and junctions that connect pipes and provide water with a defined demand. More formally, a WDN is represented as directed graphs $G=(N, E)$, consisting of the set $N$ of nodes (reservoirs, tanks and junctions) and the set $E$ of edges (the pipes), which are ordered pairs of elements of $N$. The direction of the edge is the direction of the water flows in standard conditions, from the reservoirs and tanks to the demand junctions.

\paragraph{Synthetic Dataset.}
60 synthetic Water Distribution Networks (WDNs) with different properties (total volume of water supply, number of pipes and node junctions, total length of pipes, etc.), distribution of customers/water demands and layout, representing the diverse nature of real WDSs. These benchmark networks are based on existing WDSs and are produced by the HydroGen model \cite{hydrogen}, which generates WDSs automatically according to a pre-defined algorithm with user defined settings for network size and characteristics.

\paragraph{Kentucky Dataset.}
15 WDNs provided by University of Kentucky and available online \cite{wdn-database}. The size of networks has been reduced to obtain networks with around 100 nodes. The resizing process is performed by cutting the more peripheral nodes (farthest nodes from the reservoirs).


\section{Results and Discussion}
In this section the simulation results are discussed: the resilience measures previously introduced in the Resilience Measures section are computed for every WDN and compared with their trophic coherence. A summary of the results is provided in Table \ref{table:summary_results}.
Finally, the importance of trophic coherence as a proxy for measuring the resilience of WDNs is compared with other topological and hydraulic measures.


\begin{table}[!ht]
\caption{Summary of the results. Synthetic networks are tested with static (SD) and variable (VD) demand patterns. Kentucky WDNs are tested with the pattern provided.
In the variable demand mode, the disruptions are created during the time of the day with peak water demand.}
\label{table:summary_results}
\centering
\small
\renewcommand{\arraystretch}{1.25}
 \begin{tabular}{|c|c c c|} 
 \hline
 & \multicolumn{2}{c}{Synthetic WDNs} & Kent. WDNs \\
      & SD & VD  &  \\ [0.5ex] 
     \hline 
     \hline
    Junction breakage & & &  \\
     \hline
    Mean \% junct. with deficit \textgreater 25\% & -0.66 & -0.65 & -0.55\\
    Mean \% junct. with deficit \textgreater 50\% & -0.66 & -0.66 & -0.56 \\
    Mean \% pop. impacted (deficit \textgreater 25\%) & -0.66 & -0.65 & -0.39 \\
    \% Pop. impacted (deficit \textgreater 50\%) & -0.66 & -0.66 & -0.39 \\
    Mean time to recover & 0.62 & 0.39 & 0.44\\
     \hline
    Chemical injection & & & \\
     \hline
    Mean Time to recover & 0.52 & 0.48 &  0.21 \\
    Mean \% chemical Extent & -0.29 & -0.30 & -0.52 \\
    Mean \% pop. impacted & -0.40 & -0.37 & -0.46 \\
     \hline
\end{tabular}
\normalsize
\end{table}

\subsection{WDN Resilience Results}

\subsubsection{Mean Percentage of Junctions Suffering Demand Deficit}
\begin{figure}[ht]
    \centering
    \begin{subfigure}[b]{0.48\textwidth}
      \includegraphics[width=\textwidth]{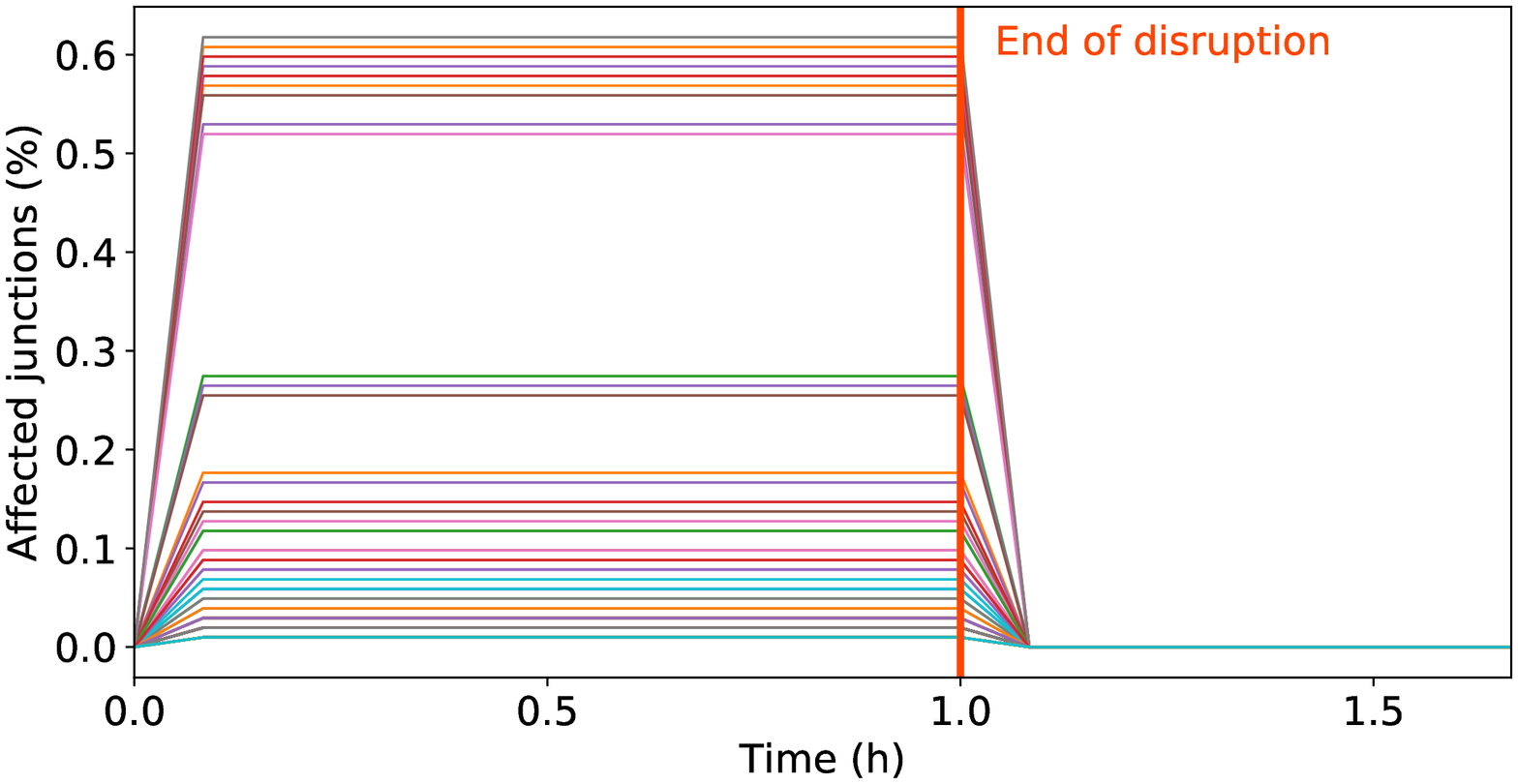}
      \caption{Number of junctions affected by demand deficit in a WDN over time, when a specific junction is closed. Each line represents a different junction closure.}
      \label{fig:demand-nodes-graph}
    \end{subfigure}
    \hfill
    \begin{subfigure}[b]{0.48\textwidth}
      \includegraphics[width=\textwidth]{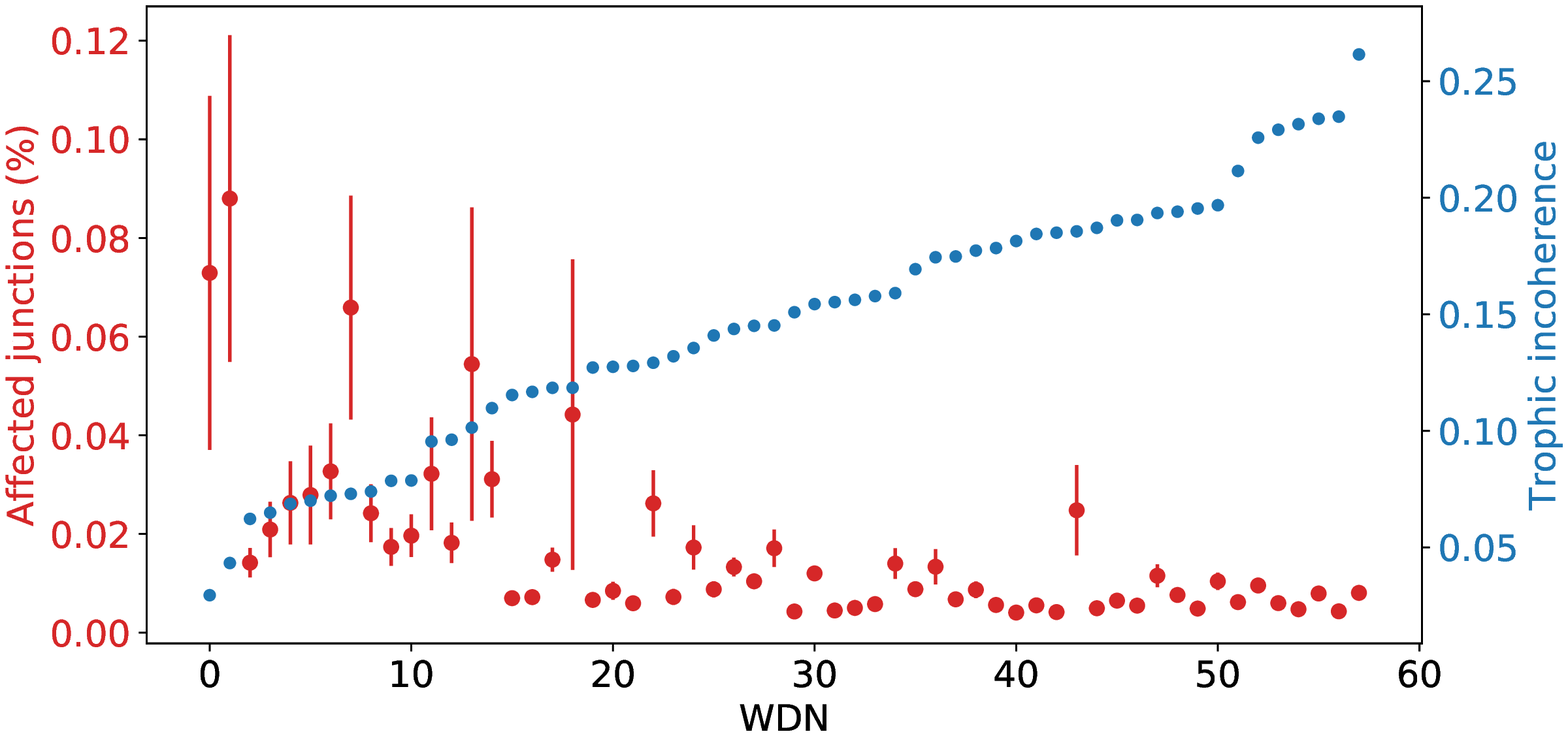}
      \caption{Comparison between mean percentage of junctions impacted by demand deficit and trophic coherence in synthetic networks.}
      \label{fig:demand-nodes}
    \end{subfigure}
    \hfill
    \caption{Water demand deficit.}
\end{figure}

The number of junctions suffering demand deficit is computed as the percentage of junctions whose supply is lower than a specified percentage of the required demand. Two thresholds have been defined: junctions with deficit higher that 25\% (minor deficit, some users may have less water than required) and junctions with deficit higher than 50\% (major deficit, most users have less water than required, some users may be without water).
An example of the percentage of junctions affected closing, in turn, different junctions is shown in Figure \ref{fig:demand-nodes-graph}: each line represents the percentage of junctions which are experiencing a disruption over time, relative to a specific junction closure. The average percentage across all demand deficit junctions is used to measure the resilience.

All the datasets show a negative correlation between percentage of junctions affected and trophic coherence (see Table \ref{table:summary_results} for details). Moreover, more coherent WDNs have higher variance.
In Figure \ref{fig:demand-nodes}, as an example, is shown the average demand deficit of all the synthetic WDNs compared with the relative trophic coherence. 

These results suggest that incoherent networks are more robust in case of possible breakages, on the contrary, coherent networks are more likely to cause widespread water demand deficits in case of a junction breakage.
This can be explained considering that the incoherence parameter measures the number of loops in a network. Loops are used to create alternative pathways for water in case of a junction closure.

\subsubsection{Mean Population Impacted by Demand Deficit}
\begin{figure}[ht]
    \centering
    \begin{subfigure}[b]{0.48\textwidth}
      \includegraphics[width=\textwidth]{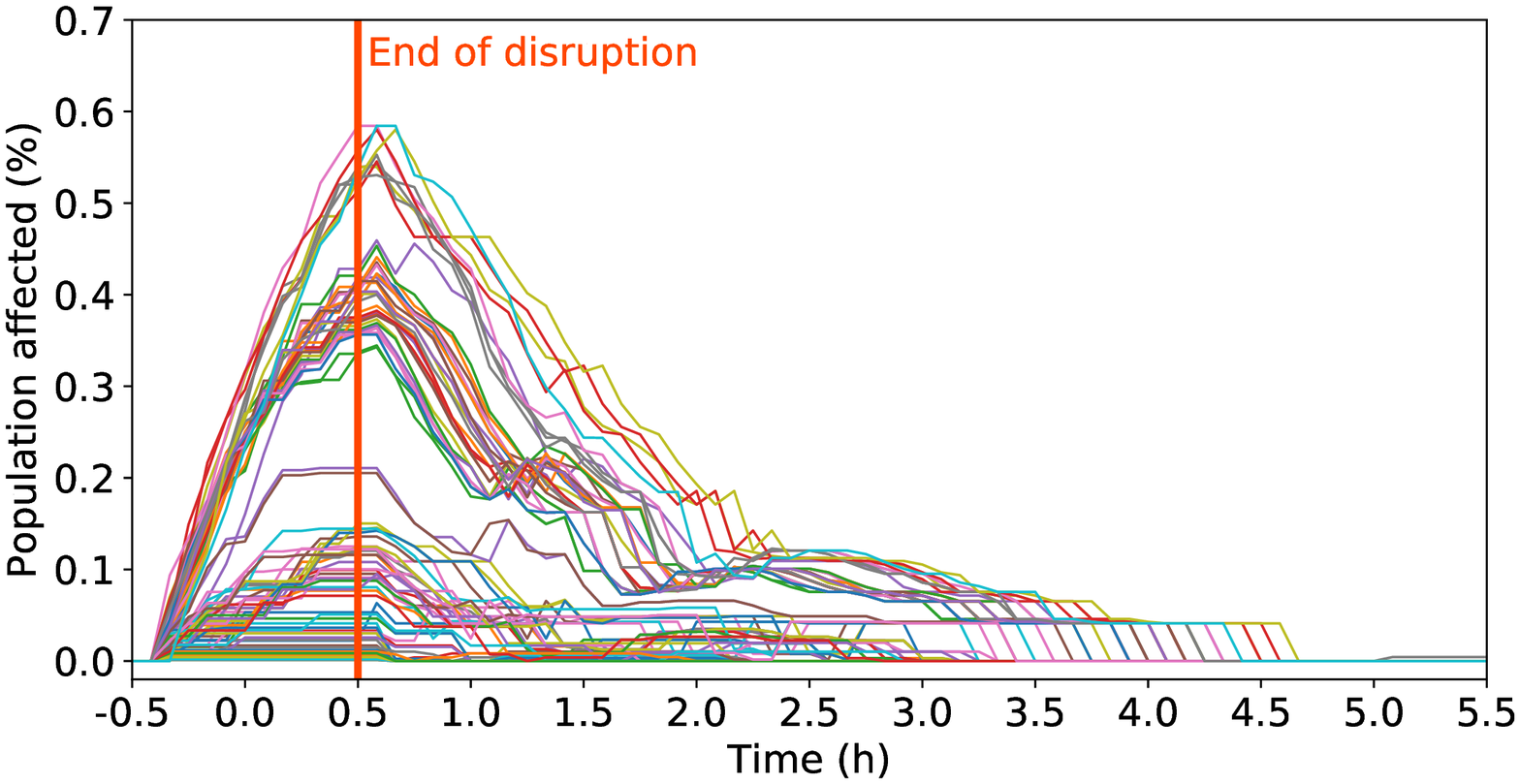}
      \caption{Population impacted by demand deficit in a WDN over time, when a specific junction is closed. Each line represents a different junction closure.}
      \label{fig:demand-population-graph}
    \end{subfigure}
    \hfill
    \begin{subfigure}[b]{0.48\textwidth}
      \includegraphics[width=\textwidth]{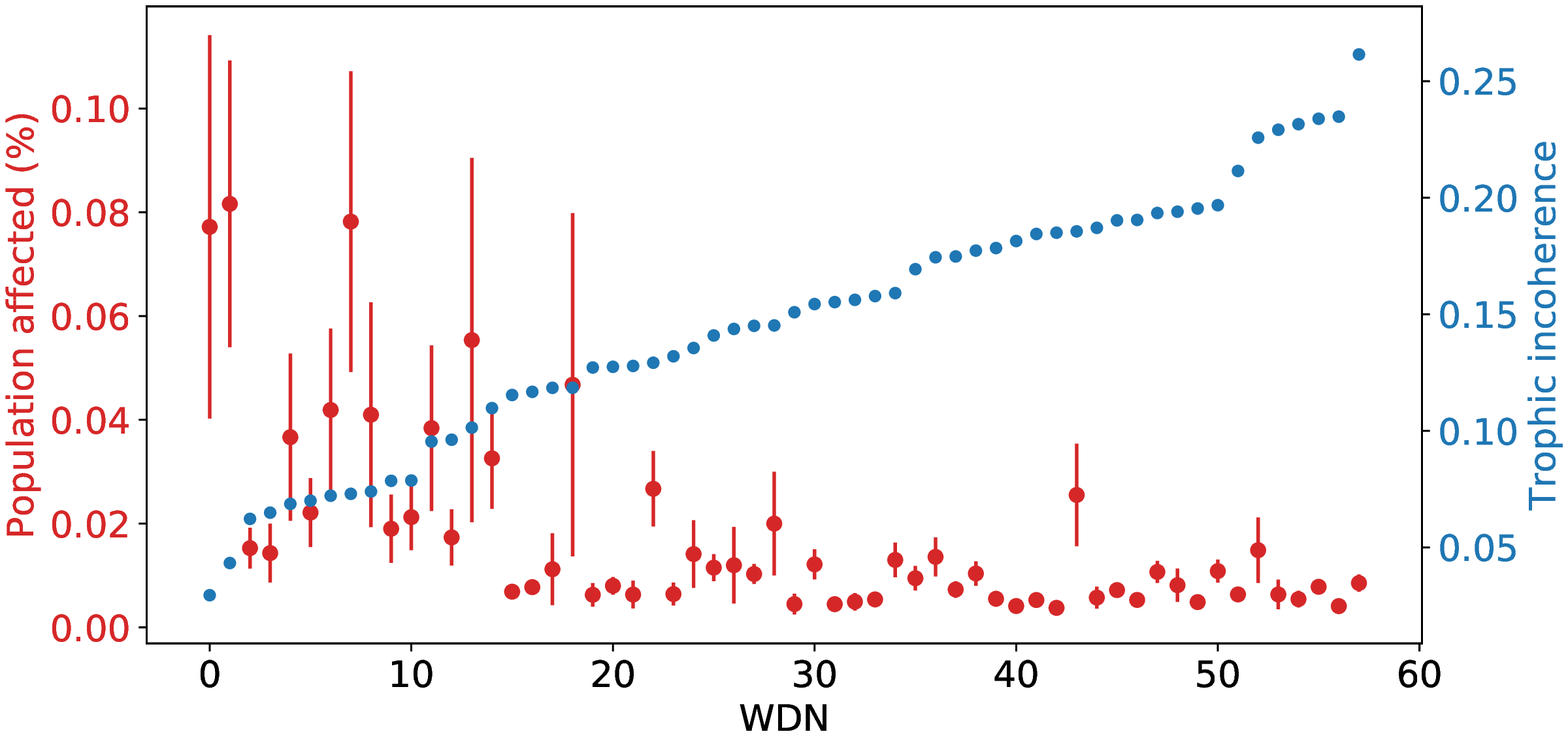}
      \caption{Comparison between mean population impacted by demand deficit and trophic coherence in synthetic networks.}
      \label{fig:demand-population}
    \end{subfigure}
    \hfill
    \caption{Population impacted by demand deficit.}
\end{figure}

The population impacted by water demand deficit in a WDN is computed as the percentage of population connected to junctions whose demand is lower than the required amount. Similar to the previous scenario, two thresholds have been chosen: population connected to junctions with deficit higher than 25\% or 50\% of the total population. An example of the percentage of population affected closing different junctions is shown in Figure \ref{fig:demand-population-graph}: each line represents the percentage of population affected by the closure of a specific junction. The average percentage is used to measure the resilience of each WDN.

All the datasets show a negative correlation between percentage of population impacted and trophic coherence (see Table \ref{table:summary_results} for details). Like for the junctions impacted, incoherence is correlated with better resilience because when a WDN contains multi-scale loops,  water can find alternative ways to reach and restore the demand in the junctions .
In Figure \ref{fig:demand-population} is shown the average percentage of population deficit compared with the trophic coherence of the synthetic WDNs. 

\subsubsection{Mean Time to Age Recovery}
\begin{figure}[ht]
    \centering
    \begin{subfigure}[hb]{0.48\textwidth}
      \includegraphics[width=\textwidth]{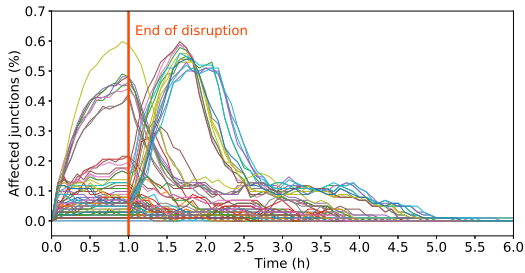}
      \caption{Age (difference with standard conditions) in a WDN over time, when a specific junction is closed. Each line represents a different junction closure.}
      \label{fig:age-time-graph}
    \end{subfigure}
    \hfill
    \begin{subfigure}[hb]{0.48\textwidth}
      \includegraphics[width=\textwidth]{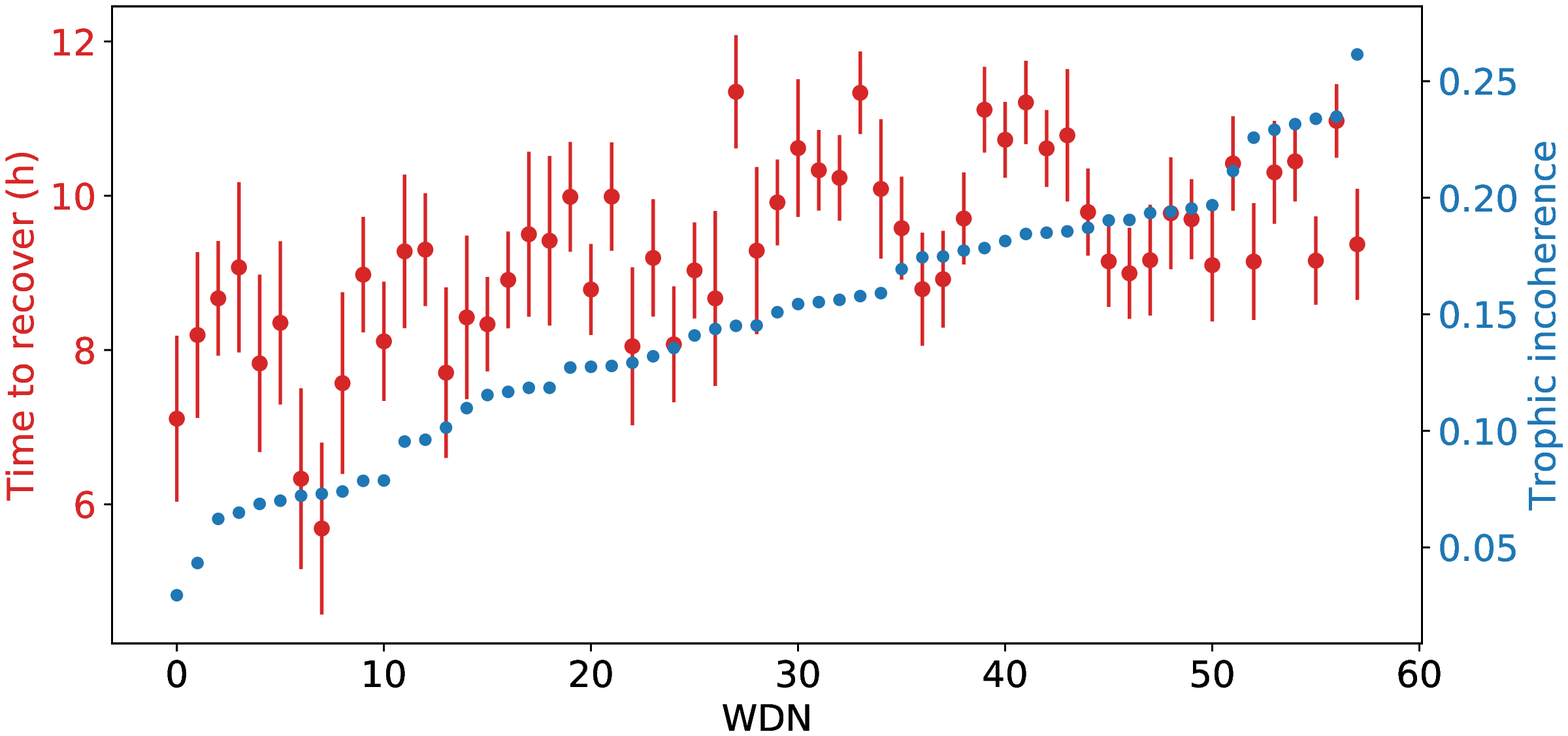}
      \caption{Comparison between mean time to age recovery and trophic coherence in synthetic networks.}
      \label{fig:age-time}
    \end{subfigure}
    \hfill
    \caption{Time to age recovery.}
\end{figure}

The time to recover the water age is computed measuring the time required to restore the standard water age in each junction. An example of how the water age varies over time, after a junction closure, is shown in Figure \ref{fig:age-time-graph}: each line is the total age increase (sum of differences in each junction with standard condition) in the junctions for a specific junction closure.

All the datasets show a positive correlation between percentage of junctions affected and trophic coherence (see Table \ref{table:summary_results} for details) suggesting that coherent networks recover faster from physical disruptions than incoherent ones. In this scenario, loops in a WDN decrease the resilience because they create water feedback loops in the pipes, increasing the recovery time. Figure \ref{fig:age-time} shows the average age recovery time compared with the trophic coherence of the synthetic WDNs.

\subsubsection{Mean Time to Chemical Contamination Recovery}
\begin{figure}[ht]
    \centering
    \begin{subfigure}[hb]{0.48\textwidth}
      \includegraphics[width=\textwidth]{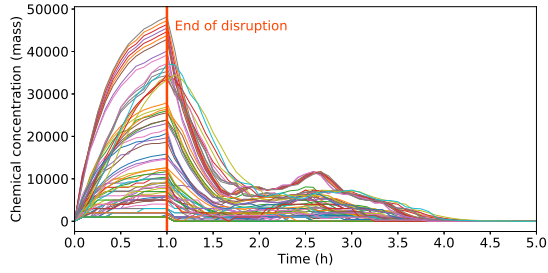}
      \caption{Total amount of chemical in a WDN over time, with injection in a specific junction. Each line represents a different injection junction.}
      \label{fig:chem-time-graph}
    \end{subfigure}
    \hfill
    \begin{subfigure}[hb]{0.48\textwidth}
      \includegraphics[width=\textwidth]{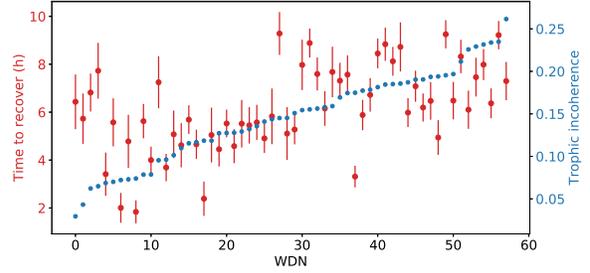}
      \caption{Comparison between mean time to recover from a chemical contamination and trophic coherence in synthetic networks.}
      \label{fig:chem-time}
    \end{subfigure}
    \hfill
    \caption{Time to chemical contamination recovery.}
\end{figure}

The mean time to chemical contamination recovery is the time required to expel the chemical from all the junctions of a WDN. An example of the amount of chemical flowing in all the junctions (sum of chemical in each junction) is shown in Figure \ref{fig:chem-time-graph}: each line represents the total amount of chemical over time, with injection in a specific junction.

Overall, a positive correlation is observed in the synthetic WDNs, Kentucky dataset has a weak positive correlation. These results confirm the previous findings in the mean time to age recovery scenario, i.e. they demonstrate that coherent networks recover faster, also from chemical contamination.
Figure \ref{fig:chem-time} shows the average expulsion time compared with the trophic coherence of the synthetic WDNs. 

\subsubsection{Mean Chemical Extent}
\begin{figure}[ht]
    \centering
    \begin{subfigure}[hb]{0.48\textwidth}
      \includegraphics[width=\textwidth]{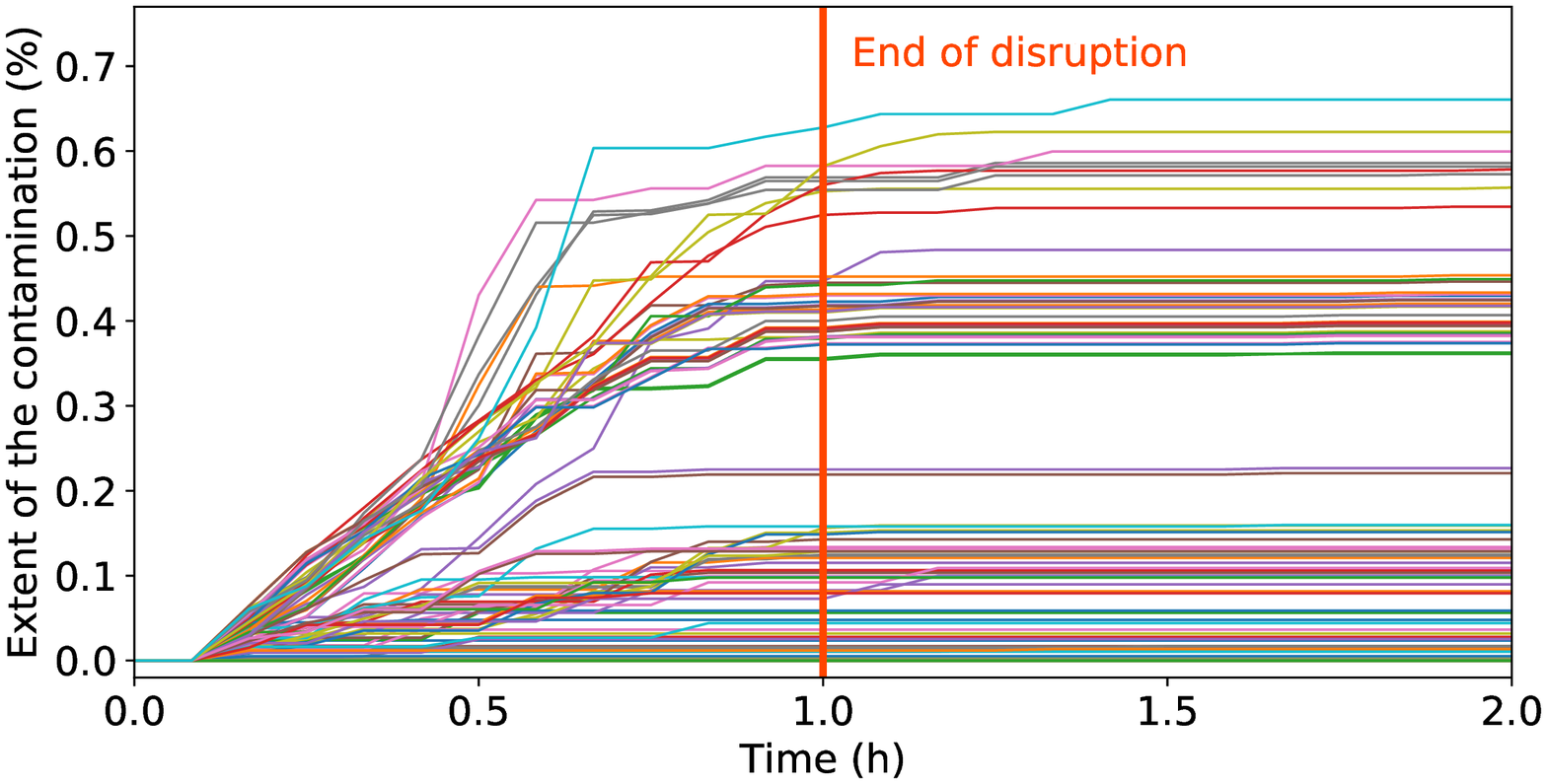}
      \caption{Extent of the chemical contamination in a WDN over time, with injection in a specific junction. Each line represents a different injection junction.}
      \label{fig:chem-extent-graph}
    \end{subfigure}
    \hfill
    \begin{subfigure}[hb]{0.48\textwidth}
      \includegraphics[width=\textwidth]{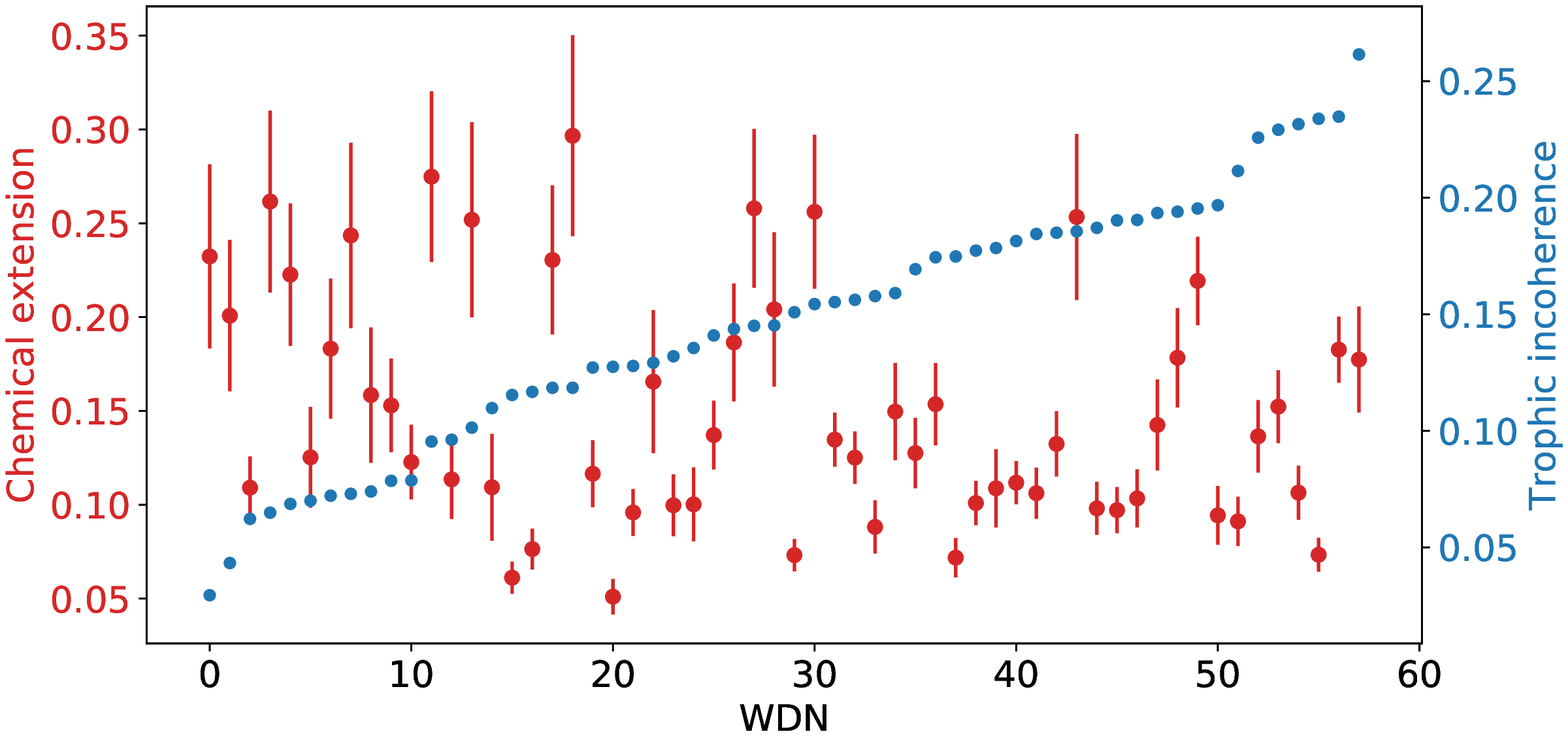}
      \caption{Comparison between mean chemical contamination extent and trophic coherence in synthetic networks.}
      \label{fig:chem-extent}
    \end{subfigure}
    \hfill
    \caption{Percentage of extent of a chemical contamination.}
\end{figure}

A chemical is injected in a selected junction and the maximum extent of the pollution in the WDN is measured in terms of total length of pipes contaminated. An example of the chemical diffusion in the pipes of a WDN is shown in Figure \ref{fig:chem-extent-graph}: each line represents the length (in percentage) of pipes affected over time, with an injection in a specific junction.

A moderate negative correlation is observed in all the datasets. These results seem to suggest that incoherent networks helps to prevent chemical from spreading in the entire network. In this scenario, loops seems to restrain the pollution. Figure \ref{fig:chem-extent} shows the average expulsion time compared with the trophic coherence of the synthetic WDNs. 

\subsubsection{Mean Population Contamination}
\begin{figure}[ht]
    \centering
    \begin{subfigure}[!ht]{0.48\textwidth}
      \includegraphics[width=\textwidth]{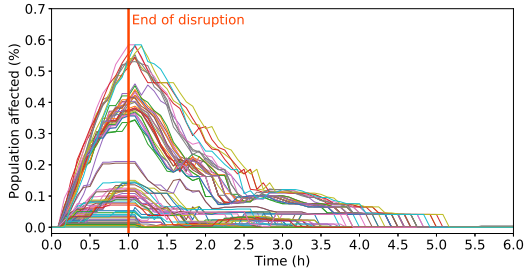}
      \caption{Percentage of population exposed to chemical in a WDN over time, with injection in a specific junction. Each line represents a different injection junction.}
      \label{fig:chem-population-graph}
    \end{subfigure}
    \hfill
    \begin{subfigure}[!ht]{0.48\textwidth}
      \includegraphics[width=\textwidth]{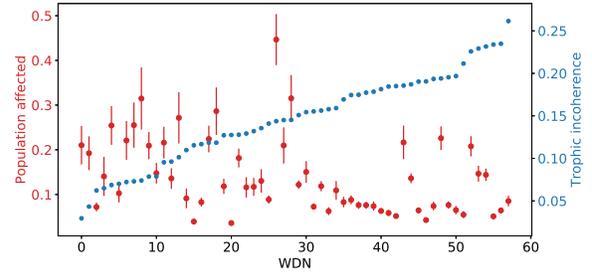}
      \caption{Comparison between mean percentage of population exposed to chemical and trophic coherence in synthetic networks.}
      \label{fig:chem-population}
    \end{subfigure}
    \hfill
    \caption{Percentage of population exposed to chemical.}
\end{figure}

The population contaminated by a chemical injection in a WDN is computed as the percentage of population connected to junctions which have been contaminated. An example of the percentage of population affected by chemical contamination in a WDN is shown in Figure \ref{fig:chem-population-graph}: each line represents the percentage of affected population over time with initial injection in a specific junction.

A moderate negative correlation is observed in all the datasets, suggesting that when a chemical is injected in a coherent network, it will affect more people than in an incoherent one. Like for the mean chemical extent, also in this scenario, loops seems to restrain the pollution.

Figure \ref{fig:chem-population} shows the percentage of contaminated population compared with the trophic coherence of the synthetic WDNs.

\subsection{Comparison with other Resilience Metrics}
A resilience ensemble is created aggregating the trophic coherence with other network topology metrics (e.g., central point dominance, mean degree) already known to be related with resilience and WDN specific properties (e.g., number of reservoirs, total water demand). This ensemble is tested by training a random decision forest with the goal of predicting the expected resilience of each network according to the metrics used in this work (i.e., number of junctions with demand deficit, mean time to recovery, and so on).

Random decision forests \cite{Ho:1995:RDF:844379.844681} work by splitting the data into subsets which most heavily belong to one class. Each tree in the forest will continue to build different subsets until it understands and represents the relationship of the variables with the target.
Random decision forests calculate their splits by mathematically determining which split will most effectively help distinguish the classes.
This information can be used to infer which metrics in the ensemble are more statistically relevant for an accurate prediction and which are unnecessary. This technique is used to classify the importance of the selected metrics. All the metrics commonly used to measure the properties of the complex networks are selected and hereafter listed: trophic coherence, number of nodes (junctions), number of edges, mean degree, link density, diameter, average shortest path length (ASPL), central point dominance, algebraic connectivity, critical ratio of defragmentation, articulation points, bridges, water sources, pipes length and total water demand.
As already stated, the objective is to detect the most important features and not to classify new networks. For this reason the normalised MSE shown in the predictive scenarios refers to the classification of the same set of data used for training, with the solely purpose to show that the model is accurate using a subset of resilience metrics.

In this work, all the metrics are computed using the Python library \textit{NetworkX} \cite{networkx-paper}, a description of the measures is provided in \cite{networkx-algorithms}. The random decision forest models are created using the Python library \textit{Scikit Learn} \cite{scikit-learn} and they are fine-tuned by using the grid search approach. 

\subsubsection{Predictive Scenarios}
The more relevant metrics for the training of random decision forest models are hereafter listed for each predictive scenario.
\noindent
\textit{Mean percentage of junctions with water demand deficit} (normalised MSE: 0.0022): pipes length (19\%), \textbf{trophic coherence} (12\%), number of edges (10\%), central point dominance (8\%), total water demand (8\%), mean degree (7\%).

\noindent
\textit{Mean percentage of population affected by demand deficit} (normalised MSE: 0.0027): pipes length (19\%), number of edges (12\%), mean degree (12\%), \textbf{trophic coherence} (10\%), bridges (7\%), central point dominance (7\%), total water demand (6\%) and articulation points (5\%).

\noindent
\textit{Mean time to age recovery} (normalised MSE: 0.0011): bridges (20\%), articulation points (18\%), mean degree (12\%), critical ratio of defragmentation (11\%), pipes length (9\%), \textbf{trophic coherence} (6\%) and number of edges (6\%).

\noindent
\textit{Mean time to chemical recovery} (normalised MSE: 0.0135): \textbf{trophic coherence} (40\%), bridges (23\%), articulation points (18\%) and total water demand (8\%).

\noindent
\textit{Mean population affected by chemical exposition} (normalised MSE: 0.0047): pipes length (15\%), total water demand (13\%), link density (10\%), \textbf{trophic coherence} (10\%), number of edges (8\%), ASPL (7\%), number of water sources (7\%), number of nodes (7\%) and algebraic connectivity (6\%).

\noindent
\textit{Mean chemical extent} (normalised MSE: 0.0149): pipes length (16\%), link density (14\%), number of edges (13\%), number of nodes (12\%), ASPL (10\%), number of water sources (10\%), diameter (5\%) and \textbf{trophic coherence} (5\%).
\\
\\
Although trophic coherence is not the most relevant metric, except for one case (mean time to chemical recovery), it is always in the list of the more influential. This means that trophic coherence helps, when coupled with other metrics, to provide a precise resilience measure for the WDNs. Moreover, and probably most importantly, while other relevant metrics are constraints (e.g., water demand) or can not be changed (e.g., length of the pipes), the coherence of the WDNs could be dynamically augmented or diminished (even over time, according to specific requirements) in order to face a specific disruption (e.g., prevent the diffusion of a pollution by increasing incoherence).
For example, the trophic level of the nodes (and, thus, the trophic coherence of the WDNs), can be changed inverting the direction of the flows in some of the pipes by opening or closing hydraulic valves or varying the amount of water provided by each reservoir or tank.

\section{Conclusions}
In this paper the correlation between resilience of WDNs and their network properties is studied using an innovative approach: data provided by simulation scenarios is used to help find a metric that measures the resilience of WDNs in different failure scenarios. Trophic coherence, which measures the stability of multi-scale feedback loops and motifs on complex networks, has been proposed as a proxy measure for resilience.

Different disruption scenarios (pipe burst and chemical contamination) are simulated and their resilience to the disruptions is correlated with the trophic coherence of each WDN. Our results show, on one hand, that trophic coherence is negatively correlated with the percentage of junctions and population affected (water deficit, chemical contamination) by a disruption. On the other hand, trophic coherence is positively correlated with the mean time to recover from a disruption. In other words, coherent networks recover faster but with a higher percentage of network junctions and populations affected (water deficit, chemical spread). Therefore, there is a trade-off between what is wanted in resilience and the coherence of the network.

Finally, trophic coherence is compared with other known resilience measures. Results show that it is always a relevant metric, even if not the most prominent, in measuring resilience in different scenarios. It is worth noting that, while other important metrics are static (e.g., length of pipes, number of junctions) or constraints (e.g., water demand), trophic coherence is a dynamic property that can change over time (e.g., opening or closing valves) according to required behaviours (e.g., reduce population impacted, rapidly expel pollution). This makes trophic coherence a valuable metric, that can also be changed over time, to improve different aspects of resilience as required in a given moment. \\

\textbf{Contributions:} A.P. conducted the simulations and analysis. F.M. and G.F. advised on the water distribution simulation, state-of-the-art and provided data. M.M. and W.G. advised on the complex network state-of-the-art. W.G. and A.P. developed the methodology for analysis. All authors contributed to writing the paper. \\

\textbf{Acknowledgements:} The authors (A.P., M.M. \& W.G.) acknowledge funding from the Lloyd's Register Foundation's Programme for Data-Centric Engineering at The Alan Turing Institute. The authors (M.M., G.F. \& W.G.) acknowledge funding from The Alan Turing Institute under the EPSRC grant EP/N510129/1. The authors (F.M. \& G.F.) acknowledge funding from EPSRC BRIM: Building Resilience Into risk Management (EP/N010329/1). \\
The authors acknowledge Microsoft Corporation for providing cloud resources on Microsoft Azure via the Alan Turing Institute.


\bibliographystyle{unsrt}  
\bibliography{bibfile}

\end{document}